\documentstyle[preprint,aps,epsf,floats]{revtex}

\begin{document}
\preprint{\tighten \vbox{\hbox{} }}
\title{Naked Singularity and Gauss-Bonnet Term in Brane World Scenarios}
\author{Ian Low$^{1,2}$ and A.\ Zee$^1$}
\address{\vspace{.7cm}$^1$Institute for Theoretical Physics,\\
University of California, Santa Barbara, CA 93106 \\
$^2$Department of Physics,\\
Carnegie Mellon University, Pittsburgh, PA 15213}
\maketitle

\begin{abstract}
We add a Gauss-Bonnet term to the Einstein-Hilbert action and study the
recent proposal to solve the cosmological constant problem. We also consider
the possibility of adding a dilaton potential to the action. In the absence
of supersymmetry, we obtain first order Bogomol'nyi equation as a
solution-generating method in our scenario. When the coefficient of the
Gauss-Bonnet term is positive, the dilaton potential is bounded below.
Assuming a simple double-well potential, we find the dilaton field to be a
kink in the fifth dimension.
\end{abstract}

{\tighten
}



\newpage

\section{Introduction}

It was suggested long time ago\cite{old} that the cosmological constant
problem could be solved if one is willing to go to higher dimensional
spacetime. Recently, a number of authors \cite{KSS,HDKS,Chen,deA} have outlined
possibilities for solving this long standing problem within the brane world
scenario \cite{n1,RS1}. In \cite{KSS,HDKS}, our 4-dimensional world is
embedded in a 5-dimensional universe and remains flat in the presence of an
arbitrary vacuum energy density $V.$ The price one pays for this remarkable
phenomenon is the appearance of naked singularities in the 5-dimensional
universe.

In this note, we consider adding to the action higher derivative terms, in
particular the Gauss-Bonnet term, and/or a potential for the dilaton field.
A simple anti-de Sitter solution and inflationary solutions to the action
with Gauss-Bonnet term were studied in \cite{Kim1,Kim2}. Models with gravity
coupled to scalars are considered in \cite{sca,DFGK}, and in the context of
five dimensional supergravity as well\cite{sup}. Here we explore the
question of how the solution of the cosmological constant problem might be
modified by the addition of these terms.

To introduce the subject and to set the stage for our discussion, let us
review the proposed solution reduced to its simplest version. We will follow
\cite{KSS} which we will refer to as KSS. The low energy effective action is
taken to be
\begin{equation}  \label{action}
S=\int d^{5}x\sqrt{-G}\left[R-\frac{4}{3}(\nabla \varphi )^{2}\right]+ \int
d^{4}x\sqrt{-g}(-V) .
\end{equation}
Gravity and a dilaton field $\varphi $ live in the 5-dimensional world and
is coupled to a thin 4-dimensional brane whose position is taken to be at $%
y=0.$ Here $M,N=0,1,2,3,5$ and $\mu ,\nu =0,1,2,3.$ The metric $g_{\mu \nu
}=\delta _{\mu }^{M}\delta _{\nu }^{N}G_{MN}(y=0)$ is the 4-dimensional
metric on the brane. (We use the convention in which $R_{NPQ}^{M}=+\Gamma
_{NQ,P}^{M}\cdots $, $R_{MN}=R_{MPN}^{P}$, $R=R_{\;M}^{M}$, and the
signature $(-,+,+,+,+).$)

Einstein's equations read
\begin{equation}
R_{MN}-\frac{1}{2}G_{MN}R-\frac{4}{3}\left[ \nabla _{M}\varphi \nabla
_{N}\varphi -\frac{1}{2}G_{MN}(\nabla \varphi )^{2}\right] +\frac{1}{2}\sqrt{%
\frac{g}{G}}\,V\,g_{\mu \nu }\,\delta _{M}^{\mu }\,\delta _{N}^{\nu
}\,\delta (y)=0  \label{eins}
\end{equation}
with $g$ and $G$ the determinant of $g_{\mu \nu }$ and $G_{MN}$
respectively. The equation of motion for the dilaton field is given by
\begin{equation}
\nabla ^{2}\varphi =0.  \label{scalar}
\end{equation}

A theorist living in the 4-dimensional brane world would notice that the
action contains a cosmological constant given by the vacuum energy density $%
V.$ KSS showed, however, that for any $V$ there exists a solution to (\ref
{eins}) and (\ref{scalar}) with the metric taking the form
\begin{equation}
ds^{2}=e^{2A(y)}(-dt^{2}+dx_{1}^{2}+dx_{2}^{2}+dx_{3}^{2})+dy^{2}.
\label{metric}
\end{equation}
In other words, the 4-dimensional world is flat in spite of the presence of $%
V.$ The dilaton field adjusts itself so that this solution exists. KSS
referred to this as a self-tuning solution of the cosmological constant
problem.

Arithmetically, this is possible because the equations of motion are solved
separately for $y>0$ and $y<0$ and we have enough arbitrary integration
constants that we can adjust in order to get the flat 4-dimensional world (%
\ref{metric}) we want. The equations of motion are, where $^{\prime}$
denotes differentiation with respect to $y$,
\begin{equation}
\frac{4}{9}(\varphi ^{\prime })^{2}+A^{\prime \prime }=-\frac{1}{6}V\delta
(y),  \label{mu55s}
\end{equation}
\begin{equation}
(A^{\prime })^{2}-\frac{1}{9}(\varphi ^{\prime })^{2}=0,  \label{55s}
\end{equation}
and
\begin{equation}
\varphi ^{\prime \prime }+4A^{\prime }\varphi ^{\prime }=0 .  \label{dils}
\end{equation}
(\ref{mu55s}) is the difference between the $\mu \mu $ and $55$ components
of Einstein's equation, (\ref{55s}) the $55$ component (which does not
receive a brane contribution proportional to $\delta (y)$), and (\ref{dils})
the dilaton equation of motion. The 5-dimensional Bianchi identity states
that only two of the three equations are independent. For example, (\ref
{mu55s}) and (\ref{55s}) imply (\ref{dils}). Solving (\ref{55s}) for $%
A^{\prime }$ and inserting into (\ref{dils}) we see immediately that $%
\varphi $ is given by the logarithm of $y.$

Thus, the solution may be chosen to be, for $y_{+}>y>0$%
\begin{equation}
\varphi (y)=-\frac{3}{4}\log (y_{+}-y)+d_{+}
\end{equation}
and
\begin{equation}
A(y)=\frac{1}{4}\log (y_{+}-y)+a_{+},  \label{apeq}
\end{equation}
and for $y_{-}<y<0$
\begin{equation}
\varphi (y)=\frac{3}{4}\log (y+y_{-})+d_{-}
\end{equation}
and
\begin{equation}
A(y)=\frac{1}{4}\log (y+y_{-})+a_{-}  \label{ameq}
\end{equation}
Here $a_{+},a_{-,}d_{+,}$ $d_{-},$ $y_{+}$ and $y_{-}$are integration
constants. The continuity of $\varphi $ and $A$ at $y=0$ determines $d_{+}$
and $a_{+}$ in terms of the other constants. Integrating (\ref{mu55s}) and (%
\ref{dils}) across $y=0$ gives the jump conditions that
\begin{equation}
A^{\prime }(y=0^{+})-A^{\prime }(y=0^{-})=\frac{1}{4}\left( -\frac{1}{y_{+}}-%
\frac{1}{y_{-}}\right) =-\frac{1}{6}V  \label{ajump}
\end{equation}
and
\begin{equation}
\varphi ^{\prime }(y=0^{+})-\varphi ^{\prime }(y=0^{-})=\frac{3}{4}\left(
\frac{1}{y_{+}}-\frac{1}{y_{-}}\right) =0  \label{phjump}
\end{equation}
These two equations merely fix $y_{+}$ and $y_{-}$ in terms of $V.$ Thus,
KSS obtained a flat space solution, type I in their classification, for any $%
V$.

One can say that, in some sense, the cosmological constant problem is solved
because we have a lot of integration constants. Of course, the framework of
embedding our universe in a larger universe is of crucial importance.

The heavy price that one pays in the KSS solution is the appearance of naked
singularities at $y_{+}$ and $y_{-}.$ Near $y_{+}$ for example, the metric
components $g_{00}=g_{ii}=e^{2A(y)}$ vanish like $\sqrt{y_{+}-y}$. Various
curvature invariants, for example the Ricci scalar $R=-20(A^{\prime
})^{2}-8A^{\prime \prime }$ diverge. These singularities are not clothed by
an event horizon. That some types of naked singularity are not acceptable is
known as cosmic censorship. Recently, Gubser\cite{gub1} has studied in
detail various singularities.

For metrics of the form $ds^{2}=e^{2A(y)}g_{\mu \nu }(x)dx^{\mu }dx^{\nu
}+dy^{2}$ we can see easily that $\sqrt{-G}R=e^{2A(y)}\sqrt{-g}%
\,R_{(4)}+\cdots $ where the 4-dimensional scalar curvature $R_{(4)}$ is
constructed out of $g_{\mu \nu }(x).$ Thus, the effective 4-dimensional
Planck mass squared is proportional to $\int dye^{2A(y)}$. In order to have
a finite 4-dimensional Planck mass, KSS were forced to choose $y_{+}$ and $%
y_{-}$ positive. They (rather arbitrarily) postulated that the universe is
cutoff at $y_{+}$ and $y_{-}$ and thus obtained a finite Planck mass squared
proportional to

\begin{equation}
\int
dy\,e^{2A(y)}=\int_{0}^{y_{+}}dy\,e^{2A(y)}+\int_{y_{-}}^{0}dy\,e^{2A(y)}.
\end{equation}
Once these integrals are thus cutoff to give finite values, 
we can obtain any desired value for the
4-dimensional Planck mass squared simply by
shifting the additive integration constant allowed by (\ref{apeq}) and (\ref
{ameq}) in the solution for $A(y).$

Notice that while changing $V$ does not appear to affect our 4-dimensional
universe, which remains flat whatever the value of $V,$ it does affect the
5-dimensional universe. In particular, it moves the naked singularities
around.

This completes a necessarily brief review of KSS to which we refer for
further details. We should perhaps mention here that KSS showed that more
generally we can write $f(\varphi )$ instead of $V$ and the same conclusion
continues to hold. This is easy to see: in (\ref{mu55s}) $V$ is replaced by $%
f(\varphi)$ and in (\ref{dils}) the term $+\frac{3}{8}\delta (y)\frac{%
\partial f}{\partial \varphi}$ is added to the right hand side. A
particularly popular choice is $f(\varphi)=Ve^{b\varphi}$ as inspired by
string theory.

At this point, it may be useful to say a few words about fine tuning and
self tuning. Traditionally, one is to write down in a field theory
Lagrangian all terms with dimension less than or equal to 4 allowed by
symmetry. The coefficients of all such terms are to be regarded as free
parameters. If we arbitrarily set one coefficient equal to another or to
zero without being able to invoke a symmetry principle, then we are said to
have fine tuned. We refer to this as the strong definition of fine tuning.
Here we are dealing with a non-renormalizable 5-dimensional action (\ref
{action}) which nobody would regard as fundamental. Presumably, this action
is to be regarded as the low energy effective action of some more
fundamental theory such as string theory. A traditionalist would say that
KSS, starting with the action (\ref{action}), has already committed fine
tuning according to the strong definition by excluding all sorts of possible
terms from (\ref{action}). We feel that it is necessary to formulate a weak
definition of fine tuning, according to which one can exclude or include any
sorts of terms from the higher dimensional action without being accused of
fine tuning. Thus, we would regard the KSS solution as not a fine tuning
solution. Rather, KSS described their solution as self tuning, in the sense
that the 4-dimensional world remains flat regardless of the value of $V.$

For example, the 5-dimensional action (\ref{action}) could perfectly well
contain the term $-\Lambda \,e^{a\varphi }$ corresponding to a bulk
cosmological constant $\Lambda $, which KSS set to $0$ in their self-tuning
solutions. A traditionalist using the strong definition of fine tuning would
definitely call setting $\Lambda $ to $0$ fine tuning; in the language of
this subject, however, this is not called fine tuning. If $\Lambda $ is not
equal to $0,$ then as KSS themselves showed, the self tuning feature of
their solution is lost. To obtain a flat 4-dimensional world, one has to
adjust $V$ to have a value determined by $\Lambda $. KSS called this fine
tuning, as we think anybody would.

\section{Action with Gauss-Bonnet term}

We ask whether we could avoid naked singularities by adding higher
derivative terms such as $R^{2}$ to the Einstein-Hilbert action. At the same
time, of course we still have the highly non-trivial constraint that $%
\int_{0}^{\infty }dy\,e^{2A(y)}$ and $\int_{-\infty }^{0}dy\,e^{2A(y)}$ have
to be finite.

We replace the bulk action in (\ref{action}) by
\begin{equation}  \label{high}
S_{bulk}=\int d^{5}x\sqrt{-G}\left[
R+aR^{2}-4\,b\,R^{MN}R_{MN}+c\,R^{MNPQ}\,R_{MNPQ}-\frac{4}{3}(\nabla \varphi
)^{2}-{\cal V}(\varphi )\right] ,
\end{equation}
We have also included a potential ${\cal V}(\varphi )$ for the dilaton field.

We will proceed in stages. We will first study the effect of the higher
derivative terms with ${\cal V}(\varphi )=0.$ Then in section III we will
include ${\cal V}(\varphi ).$

Varying $S_{bulk}$ with respect to the metric tensor we obtain the equation
of motion
\begin{eqnarray}  \label{geneom}
&&\left\{ R_{MN}-\frac{1}{2}G_{MN}R-\frac{1}{2}G_{MN}\left[
\,a\,R^{2}-4\,b\,R^{PQ}\,R_{PQ}+c\,R^{PQST}\,R_{PQST}\right] \right.
\nonumber \\
&&+2\,aR\,R_{MN}-4\,c\,R_{MP}\,R_{\;N}^{P}+2\,c\,R_{MPQS}\,R_{N}^{\;\;PQS}+4%
\,(2b-c)\,R^{PQ}\,R_{MPQN}  \nonumber \\
&&+\left.
2\,(a-b)\,G_{MN}\,R_{\;\;;P}^{\,;P}-2\,(a-2b+c)\,R_{;M;N}-4\,(b-c)\,R_{MN\;%
\;;P}^{\;\;\;\;\;\;\;;P}\right\}  \nonumber \\
&=&\;T_{MN},
\end{eqnarray}
where
\begin{equation}
T_{MN}=\frac{4}{3}\left[ \nabla _{M}\varphi \nabla _{N}\varphi -\frac{1}{2}%
G_{MN}(\nabla \varphi )^{2}\right] - \frac{1}{2}\,G_{MN}{\cal V}(\varphi ).
\end{equation}
These higher derivative terms naturally arise in the low energy effective
action of string theory\cite{string,MT}. For our purposes, we do not inquire
of their deeper origin but simply treat the action as ``phenomenological.''
(For $a=b=0$, ~(\ref{geneom}) agrees with \cite{MT}.)

We will choose $a=b=c=\lambda $ so that the added term $\lambda
(R^{2}-4R^{MN}R_{MN}+R^{MNPQ}R_{MNPQ})$ is of the Gauss-Bonnet form. One
sees from the equation of motion that all terms involving fourth derivative
vanish in this case and (\ref{geneom}) reduces to the result given
in \cite{BD}. As is well
known, the Gauss-Bonnet combination is a topological invariant in $4-$%
dimensional spacetime. In higher dimensional spacetime, it is not a
topological invariant but nevertheless has particularly attractive
properties, as discovered by Zwiebach\cite{bach} and explained by Zumino\cite
{zu}. We proceed in an exploratory spirit we believe appropriate for this
stage of development of this nascent subject and do not apologize further
for this specific choice.

With the ansatz for the metric in (\ref{metric}), we obtain
\begin{equation}
\frac{4}{9}(\varphi ^{\prime })^{2}+ \left[1-4\lambda (A^{\prime
})^{2}\right]A^{\prime \prime }=-\frac{1}{6}f(\varphi )\,\delta (y),
\label{mumu}
\end{equation}
\begin{equation}
(A^{\prime })^{2}-\frac{1}{9}(\varphi ^{\prime })^{2}-2\lambda (A^{\prime
})^{4}=0,  \label{55}
\end{equation}
and
\begin{equation}
\varphi ^{\prime \prime }+4A^{\prime }\varphi ^{\prime }=\frac{3}{8}\,
\delta(y)\, \frac{\partial f}{\partial \varphi}(\varphi) .  \label{dil}
\end{equation}
The matching condition at the location of the 3-brane, $y=0$, becomes
\begin{eqnarray}  \label{amatch}
\left.\frac83\, \varphi^{\prime}(y)\right|_{0^-}^{0^+} &=& \frac{\partial f}{%
\partial \varphi}\left(\varphi(0)\right), \\
\left.- 6 \left[ A^{\prime}(y)- \frac43\,\lambda \,A^\prime(y)^3 \right]
\right|_{0^-}^{0^+} &=& f(\varphi(0)),
\end{eqnarray}
and the continuity condition for $\varphi(y)$ and $A(y)$ at $y=0$. In the
limit $\lambda \rightarrow 0$ these equations reduce to the ones studied in
KSS of course. Again, the Bianchi identity assures us that only two out of
three bulk equations are independent. After solving two equations we can use
the third one as a convenient check.

We solve these equations in the bulk, for $y>0$ say. The scalar equation (%
\ref{dil}) gives
\begin{equation}
\varphi ^{\prime }=d\,e^{-4A(y)}
\end{equation}
with $d$ an integration constant. Inserting this into (\ref{55}) we obtain
\begin{equation}
(A^{\prime })^{2}=\frac{1}{4\lambda }\left( 1\pm \sqrt{1-\frac{8\,d^{2}}{9}%
\,\lambda \,e^{-8A}}\;\right)  \label{eq}
\end{equation}
From (\ref{eq}) we see immediately that, for $\lambda >0$, we must have $%
e^{8A}>(8d^{2}/9)\,\lambda $ and hence the 4-D Planck mass squared is
infinite unless spacetime is cut off by some singularity. To solve (\ref{eq}%
), we change variable to $\kappa (y)=e^{-4A}=1/G_{tt}\,^{2}$ and (\ref{eq})
becomes
\begin{equation}
(\kappa ^{\prime })^{2}=\frac{4\kappa ^{2}}{\lambda }\left( 1\pm \sqrt{1-%
\frac{8\,d^{2}}{9}\,\lambda \,\kappa ^{2}}\;\right) ,  \label{beq}
\end{equation}
which can be readily solved.

For $\lambda >0$, we obtain
\begin{eqnarray}
y(\kappa)&=& y_0 \pm \frac1{\sqrt{8\,\lambda}} \left[\log\left(\cot\frac{%
\theta+\pi}4\right)- \csc\frac{\theta}2\right] , \\
y(\kappa)&=& y_0 \pm \frac1{\sqrt{8\,\lambda}} \left[\log\left(\tan\frac{%
\theta}4\right)- \sec\frac{\theta}2\right],
\end{eqnarray}
where
\begin{equation}
\sin\theta=\frac{2\,d}3 \sqrt{2\lambda}\,\kappa = \frac{2\,d}3 \sqrt{2\lambda%
}\,\frac1{G_{tt}^2} ,
\end{equation}
$-\pi/2 \le \theta \le \pi/2$ and $y_0$ is an integration constant. These
four solutions correspond to the four different choices of signs in (\ref
{beq}). In order to have finite 4-D Planck mass, there are again naked
singularities in the fifth dimension.

For $\lambda = -|\lambda| <0$, only the plus sign in (\ref{beq}) is allowed
and we have
\begin{equation}
y(\kappa) = y_0 \pm \frac1{\sqrt{8\,|\lambda|}} \left[2 \tan^{-1}\left(\tanh%
\frac{\psi}4\right) - \mbox{csch}\left(\frac{\psi}2\right)\right],
\end{equation}
where
\begin{equation}
\sinh\psi = \frac{2\,d}3 \sqrt{2|\lambda|}\,\kappa .
\end{equation}
In this case, the naked singularity persists if we demand finite 4-D Planck
mass. In conclusion, while we still have the nice self tuning solution of
the cosmological constant problem as in KSS we are also stuck with the naked
singularities.

\bigskip

\section{ADDING A DILATON POTENTIAL}

\label{potential}

We next explore what happens if in the action we add a potential ${\cal V}%
(\varphi )$ for the scalar field. A potential can arise in string theory
from higher order corrections, but again in the ``phenomenological'' spirit
of this paper we do not concern ourselves with its origin. The equations of
motion are modified to
\begin{equation}
\frac{4}{9}(\varphi ^{\prime })^{2}+\left[ 1-4\lambda (A^{\prime
})^{2}\right] A^{\prime \prime }=-\frac{1}{6}f(\varphi )\,\delta (y),
\label{mumu-55'}
\end{equation}
\begin{equation}
(A^{\prime })^{2}-\frac{1}{9}(\varphi ^{\prime })^{2}-2\lambda (A^{\prime
})^{4}=-\frac{1}{12}\,{\cal V}(\varphi ),  \label{55'}
\end{equation}
and
\begin{equation}
\varphi ^{\prime \prime }+4A^{\prime }\varphi ^{\prime }=\frac{3}{8}\frac{%
\partial {\cal V}(\varphi )}{\partial \varphi }+\frac{3}{8}\delta (y)\,\frac{%
\partial f}{\partial \varphi }(\varphi )  \label{dil'}
\end{equation}
When $\lambda =0$, these equations have been studied previously and a
solution-generating method inspired by supergravity was suggested in \cite
{DFGK,ST,deB}. It is shown that, for $\lambda =0$, if ${\cal V}(\varphi )$
takes the special form
\begin{equation}
{\cal V}(\varphi )=\frac{27}{4}\left( \frac{\partial W(\varphi )}{\partial
\varphi }\right) ^{2}-12\,W(\varphi )^{2},  \label{v0}
\end{equation}
then a solution to
\begin{eqnarray}  \label{a0}
\varphi ^{\prime } &=&\frac{9}{4}\,\frac{\partial W(\varphi )}{\partial
\varphi } \\
\label{a00}
A^{\prime } &=&-W(\varphi ),
\end{eqnarray}
is also a solution to the equations of motion. Furthermore, by counting the
number of integration constants one can show that the solution space of (\ref
{v0})-(\ref{a00}) coincides with the solution space of the equations of
motion following from (\ref{mumu-55'})-(\ref{dil'}) for $\lambda =0$\cite
{DFGK}. $W(\varphi )$ is sometimes called the ``superpotential'' for obvious
reason, though no supersymmetry is involved here. Note that in this case
\begin{equation}
A^{\prime \prime }=-\frac{9}{4}\left( \frac{\partial W(\varphi )}{\partial
\varphi }\right) ^{2}\le 0.  \label{app}
\end{equation}

From (\ref{mumu-55'}) and (\ref{55'}), staying in the bulk and thus
ignoring $\delta (y)$ for now, we obtain similar first order equations for
nonzero $\lambda $:
\begin{eqnarray}  \label{var}
{\cal V}(\varphi ) &=&\left[ \frac{27}{4}\left( \frac{\partial W(\varphi )}{%
\partial \varphi }\right) ^{2}+\frac{3}{2\lambda }\right] \left( 1-4\lambda
\,W(\varphi )^{2}\right) ^{2}-\frac{3}{2\lambda }, \\
A^{\prime } &=&-W(\varphi ), \\
\label{np0}
\varphi ^{\prime } &=&\frac{9}{4}\left( 1-4\lambda \,W(\varphi )^{2}\right)
\frac{\partial W(\varphi )}{\partial \varphi }.
\end{eqnarray}
In terms of the superpotential $W(\varphi )$, the matching condition at the
location of the 3-brane, $y=0$, is now
\begin{eqnarray}
\left. 6\,\frac{\partial }{\partial \varphi }\left[ W(\varphi (y))-\frac{4}{3%
}\lambda \,W(\varphi (y))^{3}\right] \right| _{y=0^{-}}^{y=0^{+}} &=&\frac{%
\partial f}{\partial \varphi }\left( \varphi (0)\right) , \\
\left. 6\,\left[ W(\varphi (y))-\frac{4}{3}\lambda \,W(\varphi
(y))^{3}\right] \right| _{y=0^{-}}^{y=0^{+}} &=&f(\varphi (0)).
\end{eqnarray}
If, in a specific model, the 3-brane tension $f(\varphi )$ is given by $12(W-%
\frac{4}{3}\,\lambda \,W^{3})$, these jump conditions can be satisfied
identically.

It is simple to check that (\ref{dil'}) is satisfied automatically.
Generalization to $n$ scalar fields {\boldmath $\varphi$}$%
=(\varphi_1,\cdots,\varphi_n)$ is achieved by the following replacement:
\begin{eqnarray}
W(\varphi) &\to& W(\mbox{\boldmath$\varphi$}), \\
\left(\frac{\partial W(\varphi)}{\partial \varphi}\right)^2 &\to& \frac{%
\partial W(\mbox{\boldmath$\varphi$})} {\partial \mbox{\boldmath$\varphi$}}%
\cdot \frac{\partial W(\mbox{\boldmath$\varphi$})} {\partial %
\mbox{\boldmath$\varphi$}}.
\end{eqnarray}
In the limit $\lambda \to 0$, these equations reduces to (\ref{v0})-(\ref
{a00}). Note that the dilaton potential ${\cal V}(\varphi)$ is now bounded
below if $\lambda$ is positive. The second derivative of $A$ now becomes
\begin{equation}  \label{appp}
A^{\prime\prime} = - \frac94 \left(\frac{\partial W(\varphi)}{\partial
\varphi}\right)^2 \left(1 - 4\lambda\, W(\varphi)^2\right)
\end{equation}
and is no longer to be non-positive always. In five dimensional gauged
supergravity, (\ref{a0}) and (\ref{a00}) arise as conditions for unbroken
supersymmetry, and (\ref{app}) is used to prove a $c-$theorem\cite{FGPW}. It
would be interesting to study the implications of (\ref{var})-(\ref{np0}) in
the context of these models.

The reason that we are able to obtain the first order Bogomol'nyi equations
in the presence of higher derivative terms is that we have chosen the
particularly nice Gauss-Bonnet combination. For our purposes we regard the
method as simply a method for solving coupled differential equations. Thus,
for a given choice of $W$ we generate a solution for some ${\cal V}(\varphi
).$ Note that if we did not use the Gauss-Bonnet combination we would have
third and fourth derivatives of $A$ appearing in (\ref{mumu-55'}).

We now choose $W(\varphi )=s\,\varphi $ to be a linear function of $\varphi $%
. Note that $s$ has dimension of an inverse length. We find it convenient to
define the length scale $l=2/(9\,\sqrt{\lambda }\,s^{2})$ and the
dimensionless ratio $\sigma =(l/\sqrt{\lambda})^{\frac{1}{2}}.$

Following the steps outlined above, we generate the solution for $y>0$%
\begin{equation}  \label{s1}
\varphi (y)=\frac{3}{2\sqrt{2}}\,\sigma \tanh \left( \frac{y-y_{+}}{l}%
\right) ,
\end{equation}
\begin{equation}
A(y)=-\frac{1}{2}\,\sigma ^{2}\log \left[ \cosh \left( \frac{y-y_{+}}{l}%
\right) \right] +k_{+}
\end{equation}
and for $y<0$%
\begin{equation}
\varphi (y)=\varepsilon \frac{3}{2\sqrt{2}}\,\sigma \tanh \left( \frac{%
y-y_{-}}{l}\right) ,  \label{negphi}
\end{equation}
\begin{equation}  \label{s2}
A(y)=-\frac{1}{2}\,\sigma ^{2}\log \left[ \cosh \left( \frac{y-y_{-}}{l}%
\right) \right] +k_{-}
\end{equation}
The potential is a familiar double well
\begin{equation}
{\cal V}(\varphi )=\frac{3}{2\lambda }\left( \frac{1}{\sigma ^{2}}+1\right)
\left( 1-\frac{8}{9}\,\frac{\varphi ^{2}}{\sigma ^{2}}\right) ^{2}-\frac{3}{%
2\lambda },
\end{equation}
Note from (\ref{dil'}) that we have freedom in choosing the sign of $
\varphi $ represented by $\varepsilon =\pm 1$ in (\ref{negphi}). The $
\varphi $ solution we have is a kink in the fifth dimension interpolating
two vacua of the potential. The spacetime is asymptotically $AdS$ and there
is no singularity at all. Moreover, the 4-D Planck mass is finite.

Unfortunately, in this particular example, the self tuning feature of KSS is
also lost. To see this, take $f(\varphi )=V$ for simplicity. The continuity
of $\varphi $ and $\varphi ^{\prime }$ fixes $\varepsilon =+1$ and $%
y_{+}=y_{-}\equiv y_{*}.$ The continuity of $A$ fixes $k_{+}=$ $k_{-}$ while
(\ref{amatch}) tells us that the jump in $[1-\frac{4}{3}\lambda (A^{\prime
})^{2}]A^{\prime }$ across $y=0$ is equal to $-\frac{1}{6}V.$ But this
cannot be if $V$ is not zero since $A(y)=-\frac{1}{2}\,\sigma ^{2}\log
\,\cosh \frac{1}{l}(y-y_{*})$ is perfectly smooth across $y=0.$ The crucial
point here is that we no longer have the freedom of including an additive
constant in the solution for $\varphi $ in this example.

We can also choose a more general superpotential of the  form $W(\varphi
)=s\,\varphi +r$. We find that, by choosing $\lambda <0$ and $s=\sqrt{%
2/(9|\lambda |)}$, the dilaton potential is a constant
\begin{equation}
{\cal V}(\varphi )=\frac{3}{2|\lambda |}
\end{equation}
which acts like a bulk cosmological constant and is independent of $r$ in
the superpotential $W(\varphi )$. Thus $r$ plays the role of an integration
constant and we recover the self tuning feature in KSS. Unfortunately, with
negative $\lambda $ the hyperbolic functions in (\ref{s1})-(\ref{s2}) turn
into trigonometric functions and we again have naked singularities in the
bulk.

In these two examples we constructed, we need either fine tuning to avoid
the naked singularities or naked singularities to maintain the self tuning
feature.

\section*{Note Added}

After this paper was submitted a paper\cite{Cs} appeared in which the
authors proved a no-go theorem which states that, in the scenario of \cite
{KSS,HDKS}, one needs either fine-tuning or naked singularities to achieve
the flatness of our universe. Although the two examples we constructed here,
in the presence of Gauss-Bonnet term, are consistent with this no-go
theorem, we would like to point out that a crucial ingredient of the
proof given in \cite{Cs}
, $A^{\prime \prime }\le 0$, is not true in our case, as can be seen
from (\ref{appp}). Therefore there might still be hope of retaining
self-tuning feature without invoking naked singularities.

\bigskip
\acknowledgments
We thank Lisa Randall and Eva Silverstein for informative discussions and
for encouragement, and David Gross, Steve Gubser, Gary Horowitz, and Joseph
Polchinski for helpful comments. I.\,L. also benefitted from discussions
with Aki Hashimoto and Michael Quist. This work was supported in part by the
NSF under grant number PHY 89-04035 at ITP and by Department of Energy under
grant number DOE-ER-40682-143 at CMU. I.\,L. is supported in part by an ITP
Graduate Fellowship.

\tighten



\begin{references}
\bibitem{old}  V. Rubakov and M. Shaposhnikov, Phys. Lett., {\bf B125}
(1983) 139.

\bibitem{KSS}  S. Kachru, M. Schulz, and E. Silverstein, hep-th/0001206.

\bibitem{HDKS}  N. Arkani-Hamed, S. Dimopoulos, N. Kaloper and R. Sundrum,
hep-th/0001197.

\bibitem{Chen}  J.W. Chen, M.A. Luty and E. Ponton, hep-th/0003067.

\bibitem{deA}
S.P. de Alwis, hep-th/0002174; S.P. de Alwis, A.T. Flournoy and N. Irges,
hep-th/0004125.

\bibitem{n1}  N. Arkani-Hamed, S. Dimopoulos and G. Dvali, Phys. Lett. {\bf %
B429} (1998) 263; I. Antoniadis, N. Arkani-Hamed, S. Dimopoulos and G.
Dvali, Phys. Lett., {\bf B436} (1998) 257.

\bibitem{RS1}  R. Sundrum and L. Randall, Phys. Rev. Lett. {\bf 83} (1999)
3370; R. Sundrum and L. Randall, Phys. Rev. Lett. {\bf 83} (1999) 4690.

\bibitem{Kim1}  J.E. Kim, B. Kyae and H.M. Lee, hep-ph/9912344.

\bibitem{Kim2}  J.E. Kim, B. Kyae and H.M. Lee, hep-th/0004005.

\bibitem{sca}  W.D. Goldberger and M.B. Wise, Phys. Rev. {\bf D60} (1999)
107505; W.D. Goldberger and M.B. Wise, Phys. Rev. Lett. {\bf 83} (1999)
4922; N. Kaloper, Phys. Rev. {\bf D60} (1999) 123506; M. Gremm,
hep-th/9912060; C. Csaki, J. Erlich, T.J. Hollowood and Y. Shirman,
hep-th/0001033; M. Gremm, hep-th/0002040; S.B. Giddings, E. Katz and L.
Randall, hep-th/0002091.

\bibitem{DFGK}  O. DeWolfe, D.Z. Freedman, S.S. Gubser, and A. Karch,
hep-th/9909134.

\bibitem{sup}  A. Lukas, B. Ovrut, K. Stelle and D. Waldram, Phys. Rev. {\bf %
D59} (1999) 086001; K. Behrndt and M. Cvetic, hep-th/9909058; A. Chambin and
G.W. Gibbons, hep-th/9909130; C. Grojean, J. Cline and G. Servant,
hep-th/9910081; I. Bakas, A. Brandhuber and K Sfetsos, hep-th/9912132; R.
Kallosh, A. Linde and M. Shmakova, JHEP, {\bf 9911} (1999) 010.

\bibitem{gub1}  S.S. Gubser, hep-th/0002160.

\bibitem{string}  D.J. Gross and E. Witten, Nucl. Phys., {\bf B277} (1986)
1; D.J. Gross and J.H. Sloan, Nucl. Phys., {\bf B291} (1987) 41; R.R.
Metsaev and A.A. Tseytlin, Phys. Lett., {\bf B191} (1987) 354; R.C. Meyers,
Phys. Rev. {\bf D36} (1987) 392; C.G. Callan, R.C. Meyers, and M.J. Perry,
Nucl. Phys. {\bf B311} (1988) 673.

\bibitem{MT}  R.R. Metsaev and A.A. Tseytlin, Nucl. Phys., {\bf B293} (1987)
385.

\bibitem{BD}  D.G. Boulware and S. Deser, Phys. Rev. Lett. {\bf 55} (1985)
2656.

\bibitem{bach}  B. Zwiebach, Phys. Lett., {\bf B156} (1985) 315.

\bibitem{zu}  B. Zumino, Phys. Rept., {\bf 137} (1986) 109.

\bibitem{ST}  K. Skenderis and P.K. Townsend, Phys. Lett., {\bf B468} (1999)
46.

\bibitem{deB}  J. de Boer, E. Verlinde and H. Verlinde, hep-th/9912012.

\bibitem{FGPW}  D.Z. Freedman, S.S. Gubser, K. Plich, and N.P. Warner,
hep-th/9904017.

\bibitem{Cs}  C. Csaki, J. Erlich, C. Grojean and T. Hollowood, hep-th/0004133.
\end{references}
\end{document}